\documentclass[lettersize,journal]{IEEEtran}

\input{packages}

\begin{document}


\title{\sysname: Virtualizing Arm CCA with TrustZone}

\author{Xiangyi Xu, Wenhao Wang\textsuperscript{\Envelope}, Yongzheng Wu, Chenyu Wang, Huifeng Zhu, Haocheng Ma, \\ Benshan Mei, Zixuan Pang, Rui Hou, and Yier Jin\textsuperscript{\Envelope}
\IEEEcompsocitemizethanks{
\IEEEcompsocthanksitem Corresponding authors: Wenhao Wang (\href{mailto:wangwenhao@iie.ac.cn}{wangwenhao@iie.ac.cn}) and Yier Jin (\href{mailto:jinyier@ustc.edu.cn}{jinyier@ustc.edu.cn}).
\IEEEcompsocthanksitem X. Xu, Y. Wu, C. Wang, H. Zhu and H. Ma are with Huawei Technologies.
\IEEEcompsocthanksitem W. Wang and R. Hou are with the State Key Laboratory of Cyberspace Security Defense, Institute of Information Engineering, Chinese Academy of Sciences, and University of Chinese Academy of Sciences.
\IEEEcompsocthanksitem B.Mei is with Shanghai Jiao Tong University. Work performed while at University of Chinese Academy of Sciences.
\IEEEcompsocthanksitem Z. Pang and Y. Jin are with University of Science and Technology of China.
\IEEEcompsocthanksitem The authors from Institute of Information Engineering were supported by the National Natural Science Foundation of China (Grant No. 62272452), the Strategic Priority Research Program of the Chinese Academy of Sciences (Grant No. XDB0690100) and the research grant from Huawei Technologies.
}}

\maketitle 

\begin{abstract}
Arm introduced the Confidential Compute Architecture (CCA) in the upcoming Armv9-A architecture, enabling the support of confidential virtual machines ({\cVM}s) in a separate world called the Realm world, providing protection from untrusted normal world. While CCA offers a promising future for confidential computing, the widespread commercial CCA hardware is not available in the near future. To fill this gap, we present \textit{\sysname}, an architecture that facilitates virtualized CCA using TrustZone, a mature hardware feature on existing Arm platforms. 
Notably, \sysname can be implemented on platforms equipped with the Secure EL2 (S-EL2) extension from ARMv8.4 onwards, as well as on earlier platforms that lack S-EL2 support.
\sysname provides strong compatibility with the CCA specifications at the API level. We developed the entire CCA software and firmware stack on top of \sysname, including the enhancements to the normal world's KVM to support {\cVM}s, and the TrustZone Management Monitor (TMM) that enforces isolation among {\cVM}s and provides \cVM lifecycle management.
We have implemented \sysname on real Arm servers, both with and without S-EL2 support. Our evaluation on micro-benchmarks and macro-benchmarks demonstrates that the overhead of running {\cVM}s is acceptable compared to running normal-world VMs. Specifically, in a set of real-world workloads, the overhead of \sysname-SEL2 is less than 29.7\% for I/O intensive workloads, while \sysname-EL3 outperforms the baseline in most cases.

\end{abstract}





\section{Introduction}
\label{sec:intro}

Confidential computing is rapidly emerging as an indispensable technology in the realm of cloud computing. Its primary objective is to safeguard the sensitive information of tenants from potential risks posed by untrustworthy or improperly configured cloud service providers (CSPs). This paradigm shift towards confidential computing not only enhances data privacy and security but also instills a greater sense of trust in cloud-based services. As a result, it is increasingly becoming an integral part of modern cloud architectures.
Recognizing the significance of confidential computing, leading chip companies have stepped forward to offer support for trusted execution environments (TEEs) within their product offerings. For instance, Intel's SGX~\cite{sgxoverview} and TDX~\cite{tdx2020}, AMD's SEV~\cite{sev2020strengthening}, and IBM's PEF~\cite{hunt2021confidential} are all dedicated features that enable the creation of isolated regions (i.e., enclaves) and confidential virtual machines (i.e., \cvms) exclusively used by tenants. 

Arm processors have gained widespread popularity in mobile devices like smartphones due to their energy efficiency. Notably, Arm was one of the pioneers in supporting TEEs, i.e., TrustZone, to ensure the protection of sensitive data, such as passwords and fingerprints. TrustZone divides the system into two distinct worlds: the normal world and the secure world. While the normal world runs the feature-rich operating system (OS), the secure world handles critical operations that require enhanced security measures. To cater to the specific requirements of the secure world, customized trusted OSes like OPTEE~\cite{optee} and iTrustee~\cite{itrustee} have been developed. These tailored trusted OSes ensure the provision of secure and reliable services within the secure world. However, unlike AMD SEV, TrustZone does not fully support the cloud computing scenario, particularly in supporting a full-featured OS like Linux running inside \cvms, while the lifecycle of \cvms is managed by the host hypervisor (e.g., KVM).

Recently, Arm has made significant strides in the cloud computing market. To meet the confidential computing need, Arm has announced the CCA, a series of hardware and software architecture innovations available as part of the ARMv9-A architecture. CCA supports the dynamic creation and management of \textit{Realms} (Arm's terminology for \cvms), opening confidential computing to developers and various workloads. 


\para{Motivations}
As Arm's latest innovation for confidential computing, CCA points to a convincing future for confidential computing. However, it typically takes several years before commodity chips incorporating this technology are manufactured and made available to cloud computing companies.
In the meantime, a significant number of legacy Arm servers have already been deployed, such as AWS Graviton and Google's Tau T2A machine series. Even if CCA hardware becomes available in the future, these legacy servers are likely to remain in use for an extended period. Furthermore, there may continue to be new hardware released that does not include CCA. While these servers offer improved computational capabilities, they do not meet the requirements of confidential cloud computing. Moreover, they cannot take advantage of the evolving software support developed for CCA, such as the \cVM guest kernel, KVM, and the \textit{Realm Management Monitor} (\RMM)~\cite{rmmsamsung,ccaguest,ccakvm}, which have undertaken tremendous development efforts and innovations.


Considering the lagging behind of CCA hardware and the urgent need for confidential computing, we propose \sysname in this paper, a \textit{virtualized CCA} design on existing hardware that provides both strong isolation and good performance, and offers \textit{strong compatibility with CCA}. 
To stimulate real-world deployment, \sysname is designed to support native OS distributions (such as booting Ubuntu images) with only minimal modifications within the \cVM, so that existing applications, toolchains, and peripheral device drivers (e.g., GPU driver) can run securely with no portability effort. \footnote{As summarized in Table~\ref{table:loc}, about 1.3K lines of code are added to the KVM, and 1.3K (resp. 2.1K) lines of code are added to the \sysname-SEL2 (resp. \sysname-EL3) guest OS.}

\para{Our design}
We utilize the secure world on TrustZone as the hosting environment for the \cvms, while considering the hypervisor as \textit{entirely untrusted}. 
This contrasts with existing designs of hypervisor-based TEEs, like SeKVM~\cite{li2021secure} and protected KVM (pKVM)~\cite{pkvm}, which run the \cVM in the normal world and rely on a trusted core of the hypervisor to safeguard the confidentiality and integrity of \cVM data from the untrusted host Linux kernel. 
We argue that running \cvms in the secure world offers several advantages. Firstly, even though the trusted core in SeKVM is formally verified, it still presents challenges when working with constantly evolving commodity hypervisors, particularly when updates to the hypervisor are required. Secondly, running in the secure world allows us to utilize hardware support, including not only the TrustZone Address Space Controller (TZASC) for memory isolation but also enhanced security controls such as the TrustZone Protection Controller (TZPC) for device assignment. In contrast, hypervisor-based TEEs can only rely on the MMU for memory isolation.

Notably, \sysname stands out for its versatility in deployment on legacy hardware, regardless of Secure EL2 (S-EL2) support, unlike the existing design Twinvisor~\cite{li2021twinvisor}, which is limited to S-EL2-compatible hardware. To our knowledge, \sysname is the first CVM solution capable of running secure KVM guests in the secure world even without S-EL2 support. It is important to note that currently, there are no publicly available hardware options that offer S-EL2 support. Furthermore, \sysname prioritizes strong compatibility with CCA, while Twinvisor does not.


However, constructing CVMs within the secure world is not a straightforward task due to significant differences between the \sysname model and TrustZone. 
Within TrustZone, the trusted OS and trusted applications (TAs) operate in a separate world, managing hardware resources such as secure memory and the interrupt controller independently. In contrast, within the \sysname framework, these resources, as well as the life cycle of \cvms, are governed by the hypervisor operating in the normal world.
To alleviate trust in the hypervisor and reduce the trusted computing base (TCB), we propose a small and trusted software layer, referred to as the \textit{TrustZone Management Monitor} (TMM) -- an analogy to the RMM in CCA -- to act as a gatekeeper between the two worlds. The TMM manages CPU state transitions between the untrusted normal world and the \cvms. 

Similar to CCA, the untrusted hypervisor is still responsible for the resource and life cycle management of \cVM. Our design provides a TrustZone Management Interface (TMI) and a TrustZone Service Interface (TMI) for the interactions between the hypervisor and \cVM.
These interfaces are implemented as Secure Monitor Calls (SMCs) to the Arm Trusted Firmware (ATF) and are compatible with the Realm Management Interface (\RMI) and Realm Service Interface (RSI) at the application programming interface (API) level; that is, the input parameters, functionalities, return values, and handling of errors and exceptions are all consistent with the CCA specifications. 
Thanks to such strong compatibility, the software stack for \sysname runs on CCA with minimal changes.

Furthermore, the TMM assists the hypervisor in managing faults and exceptions originating from {\cVM}s, such as page faults and memory-mapped IO (MMIO) exceptions. 
The \TMM is also responsible for translating interrupts between the two worlds.
Specifically, with modern Generic Interrupt Controller (GIC) hardware, interrupts are assigned to different groups depending on the current secure state. To minimize the modification of guest OSes, the \TMM converts the interrupt types before switching to the host hypervisor for emulating the interrupts. Currently, \sysname supports efficient I/O virtualization for \cvms with device emulation by the host under the \texttt{virtio} IO virtualization framework.

\para{Challenge 1: Supporting Arm platforms both with and without S-EL2 support}
On Arm v8.4 and later platforms, Arm introduced support for \textit{secure virtualization}, allowing the execution of EL2 code in the secure world, known as S-EL2. On these platforms, the TMM operates at S-EL2 and serves as a translation layer between the untrusted host and the \cvms. The TMM manages all secure memory for memory virtualization, and we utilize the S-EL2 extension to ensure isolation among mutually distrusted \cvms by configuring the stage-2 page tables of the \cvms.
Regarding the virtualization of interrupts and peripheral devices, including the Power State Coordination Interface (PSCI) and timer, accesses to these devices (such as through MMIO) are initially intercepted by the TMM. Subsequently, the TMM encapsulates the request information and forwards it to the hypervisor. After the hypervisor successfully handles the requests, the TMM can resume the CVM using dedicated instructions, such as ERET.

However, earlier platforms without S-EL2 lack the essential capabilities to virtualize memory, interrupts, and devices within the secure world. 
For example, the GIC list registers, which are responsible for tracking the interrupt state and injecting vIRQ to CVM, are not supported in the secure world. Additionally, the Exception Syndrome Register (ESR), which contains the MMIO access information, is also not supported.
To enable the \cVM model, we position the TMM at EL3 as a component of the ATF, and virtualization is fully emulated through software. Specifically, we store the register state for each vCPU in the TrustZone Execution Context (TEC) associated with every \cVM and model the state transition based on the type of interrupts. The injection of vIRQ to CVM is entirely emulated within the TMM. In cases where the guest OS needs to access list registers, we replace them with SMC calls. Furthermore, we modify the guest OS by replacing MMIO accesses with SMC instructions that trap to the TMM. As TrustZone only supports a single secure world, we need to protect the \cVM from each other.
On Arm platforms lacking S-EL2 support, we prevent the guest OS from mapping the physical memory of other \cvms into its own address space by configuring write protection on the \cVM's page tables.

\para{Challenge 2: Efficient cross-world memory isolation under restricted TZASC hardware}
Existing VM-based TEEs adopt the design strategy called \textit{asynchronous management of secure memory} -- the hypervisor manages the memory allocated to every \cVM, but the allocation and recycling of these secure memory must involve and be approved by trusted components to guarantee memory isolation. 
As a result of such a strategy, since the hypervisor manages the entire physical memory, the normal and secure memory are \textit{interleaved with each other}, attaining a high memory utilization rate. 
To achieve this capability in TrustZone, the secure memory can be configured with TZASC.
However, \TZASC only supports configuring the security attributes for up to eight different memory regions, making the memory isolation while maintaining efficient memory utilization a non-trivial task.
{Twinvisor}~\cite{li2021twinvisor} tackles the challenge by introducing the split contiguous memory allocator (split CMA), such that the normal and secure worlds collaborate to resize the secure memory regions dynamically, which contributes to more than half of Twinvisor's \TCB. The dynamic mappings of secure memory to \cvms caused by memory delegation and split CMA introduce additional context switches and TLB flushing, making them inferior in memory and I/O intensive scenarios.


To address this challenge, we propose a design where the TMM takes full responsibility for managing the secure memory used by \cvms. During system initialization, a fixed and contiguous block of memory (the size of which is configurable in the BIOS) is reserved specifically for \cvms' use. When the hypervisor needs to allocate memory for a CVM, such as when loading the CVM's guest image, it triggers a TMI request. The TMM then allocates the required amount of memory and copies it from normal memory to the designated memory space for the CVM. \sysname also supports efficient allocation of various memory sizes through a single TMI request, minimizing the number of TMIs invoked.
To facilitate efficient management of secure memory, we implemented a simplified version of Linux's buddy memory system and slab algorithm within the TMM. Additionally, \sysname takes into consideration the Non-Uniform Memory Access (NUMA) affinity to improve performance. The memory allocator prioritizes the allocation of memory within the NUMA node associated with the physical CPUs assigned to the CVM. This leads to improved memory access performance (over 66.7\% improvement, as shown in Figure~\ref{fig:fileop}).

\para{Implementations} 
We implemented \sysname on real Arm servers, with and without S-EL2 support, respectively. With \sysname, we were able to build a software stack that supports both CCA and \sysname, including components such as \ATF, \RMM, QEMU/KVM, and the \test framework.
We conducted micro-benchmarks to evaluate the impact of \sysname on common hypervisor operations. Additionally, we ported a number of real-world applications, including Redis, Memcached, MongoDB, MySQL, Apache, and Nginx etc., to run on \sysname. The experimental results demonstrate that the overhead introduced by \sysname is minimal. Specifically, in a set of real-world workloads, the overhead is below 29.7\% for \sysname-SEL2, while \sysname-EL3 outperforms the baseline in the majority of workloads. This is likely attributed to the fact that in \sysname-EL3, \cvms operate in bare-metal mode, thereby eliminating the overhead of CPU and memory virtualization. This showcases the efficiency and effectiveness of \sysname in practical scenarios.

\para{Contributions} The paper makes the following contributions.

\begin{packeditemize}
    \item A virtualized CCA design atop existing TrustZone hardware, providing strong compatibility with the CCA specifications. 
    As far as we know, this is the first time that \cVM can be supported on Arm TrustZone without S-EL2 support.
    \item An entire software and firmware stack to support \sysname, including the \ATF, \RMM, host hypervisor etc. 
    \item Implementations and evaluations. 
    We have implemented our design on real Arm platforms with and without S-EL2 support respectively. The evaluations on real world benchmarks show that the design is practical.
\end{packeditemize}


\section{Background}
\label{sec:background}

\subsection{Trusted Execution Environment}
\label{subsec:tee}

In traditional outsourced computation scenario, the privileged software (such as the hypervisor) is allowed to access the client's computation. \TEE is introduced to provide a hardware-based secure partition (such as an SGX enclave or SEV VM) that ensures the integrity and confidentiality of sensitive code and data, which can only be accessed by authorized entities, even though the untrusted privileged software is still in charge of managing the available resources (such as CPU and memory). TEEs are commonly used in applications such as cloud computing, mobile payments, digital rights management, and secure browsing, and are supported by all major CPU vendors.

Existing \TEE designs mainly focus on the following aspects: \circled{1}protection of the CPU states during the transitions between the secure mode and non-secure mode, e.g., during entry to or exit from the secure mode, and during interrupts and exceptions; \circled{2}protection of the secure memory used by the secure mode. The secure memory allocated to one client cannot be mapped to other clients or the hypervisor. Optionally, the \TEE can provide memory protection against physical attacks with memory encryption; \circled{3}remote verification of the state of all \TCB components through remote attestation. During the creation of a secure instance for the client, the trusted components generate a measurement of the loaded code and data; then the measurement is signed with the hardware-backed private key, and the signed report is sent to the remote client for verification.

\para{VM-based TEEs} 
In contrast to Intel SGX, which provides isolated regions (called enclaves) within the application's address space, most recent \TEE designs (such as AMD SEV~\cite{kaplan2016amd,sev2020strengthening}, IBM PEF~\cite{hunt2021confidential} and Arm CCA~\cite{armcca}) provide the abstraction of a \VM for the TEEs. In particular, a fully-fledged guest \OS, e.g., common Linux distributions such as Ubuntu, can run securely as part of the \TEE software to support unmodified toolchains and applications, such as language runtimes, deep learning frameworks, and device drivers. VM-based \TEE embraces the cloud computing need and sets a trend for the future of confidential cloud computing. To support the \TEE functionalities, the host software (QEMU and KVM) and the guest \OS cooperates in a para-virtualization manner for the secure management of VM resources, such as the allocation and recycling of secure memory. Currently, the support for AMD SEV, Intel TDX, and Arm CCA in both Linux and QEMU/KVM is under active development. 

\subsection{Arm Platforms}
\label{subsec:arm}

\ignore{
\begin{figure}
\centering
\includegraphics[width=\columnwidth]{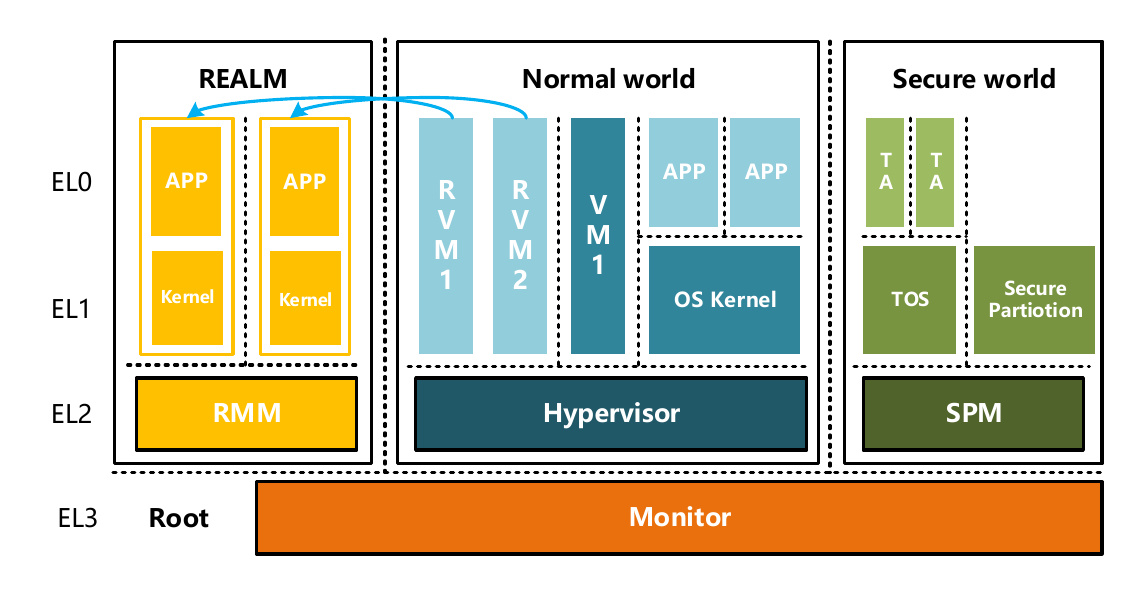}
\caption{Basics of the Arm architecture.} 
\label{fig:arm_basics}
\end{figure}
}

In the Arm architecture, the privilege levels are referred to as exception levels (ELs), since the current privilege level (CPL) can only change when the processor takes or returns from an exception. Typically, EL0 and EL1 are used by the user-level applications and by the \OS, respectively, while EL2 runs the normal hypervisor, such as KVM. The most privileged software runs at EL3, and is referred to as the \ATF.

\para{Arm TrustZone}
Analogous to the normal world for running the rich \OS and applications, TrustZone replicates the system resources in a separate world known as the secure world. A tailored trusted \OS runs at secure EL1 (S-EL1), while trusted applications (TAs) run at secure EL0 (S-EL0). The Non-Secure (NS) bit determines whether the CPU executes in the normal world or in the secure world context to create a separation in memory. The CPU state transitions between the normal world and the secure world can be triggered by the \SMC instruction. This causes a trap to the \ATF that handles the state transitions.

The separation of the secure and normal worlds protects specific memory ranges and peripherals that are only accessible by the secure world. 
The memory partitions between the normal and secure worlds are enabled by the \TZASC, an AMBA-bus compliant System-on-Chip (SoC) peripheral. With the current \TZASC hardware -- TZC-400, it is possible to define up to 8 DRAM regions as secure or non-secure. On the other hand, the \TZPC can define the peripheral as either secure or non-secure. These controllers provide secure I/O to peripherals. For instance, the \TZPC routes SPI access to the secure world, and the NS bit secures on-chip peripherals from being accessed by the normal world. 

In real-world applications, there can be multiple security services that wish to make use of \SELone. To support the feature, Arm adds the capability of running EL2 code in the secure world (known as S-EL2) in the ARMv8.4-A architecture~\cite{sel2}.
Together with \ATF, this enables multiple security services to exist in the secure world alongside each another, while the Secure Partition Manager (SPM) running at S-EL2 is responsible for managing resource partitions. 

\para{Arm virtualization extension} 
Arm introduces hardware support for efficient virtualization. For memory virtualization, Arm enables nested paging by extending the \MMU with a second stage of address translations. For CPU virtualization, Arm adds support for saving and restoring the architectural state upon context switches. 

To support secure device pass-through for Direct Memory Access (DMA) devices, similar to memory virtualization, Arm's System Memory Management Units (SMMU) support two stage address translation. For interrupt virtualization, the \GIC architecture splits logically into a Distributor block and one or more CPU interface blocks, and the virtualization extension adds a \vCPU interface for each processor in the system to the \GIC. When the \GIC sends an interrupt request (IRQ) to the processor, the interrupt is routed to the hypervisor. The hypervisor determines whether the interrupt is for itself or for a Guest \OS. 

Furthermore, the \GIC Security Extensions support the \GIC registers to take account of Secure and Non-secure accesses, and the \GIC interrupt grouping feature to support the handling of both Secure and Non-secure interrupts. In this paper, we limit the discussion on GICv3, in which Group 0 interrupts are Secure interrupts, and Group 1 interrupts can be configured as Non-secure interrupts or Secure interrupts. Moreover, a processor in Non-secure state can make only Non-secure accesses to a \GIC, while a processor in Secure state can make both Secure and Non-secure accesses to a \GIC. Considering backward compatibility for Arm \GIC, our design can also work for newer \GIC versions.

\subsection{Arm's Confidential Compute Architecture}
In response to the confidential computing need in the cloud, Arm introduced the \CCA with the Realm Management Extension (RME), which supports a new type of attestable isolation environment called a Realm -- a type of \cvms, democratizing confidential computing for all developers for the secure processing of sensitive data.
CCA consists of both hardware changes and firmware/software support. 

In particular, the firmware code in EL3 maintains the Granule Protection Table (GPT) to determine the accessibility of physical pages, and the memory management unit (MMU) is augmented with additional Granule Protection Checks (GPC) beyond traditional page table based memory protection mechanisms.
A tiny and \textit{trusted} software layer called \textit{Realm Management Monitor (RMM)} running at EL2 in the Realm world is responsible for memory isolation among Realms. The normal world hypervisor still manages the life cycle (allocation/de-allocation) of physical memory for the Realms. However, those memory can be used by Realms for various purposes only after they are explicitly delegated to the Realm through the \RMI provided by \RMM. 
The \RMI interfaces are implemented through the \SMC instruction.
When the hypervisor invokes the command, the CPU traps to EL3, which in turn switches execution to \RMM in the Realm world to handle the command. Upon completion of the \RMI interface, \RMM issues an \SMC to EL3, which switches execution back to the hypervisor in the normal world.




\subsection{Threat Model}
\label{subsec:threatmodel}


We assume a scenario where the attacker does not have physical access to the machine, which means they are unable to carry out physical attacks like cold boot or bus snooping attacks.
We trust the underlying hardware, including the Arm CPU and DRAM, and the code running at S-EL2 (the \TMM) and the \ATF. We assume the platform is equipped with reliable root of trust (RoT) and provides authentic random sources, e.g., by firmware or hardware Trusted Platform Module (TPM). Similar to existing TEEs, we assume the attackers possess the following capabilities.
\begin{packeditemize}
    \item \textit{Software attacks}. The attacker lacks physical access to the platform but may have compromised the privileged system software, namely the hypervisor, and attempts to compromise the confidentiality and integrity of the \cvms. The attacker may attempt to access secure memory by exploiting DMA.
    \item \textit{Malicious \cvms}. The attacker may construct a malicious \cVM, which tries to interfere with the execution of other \cvms, or to break the isolation enforced by the privileged system software~\cite{schwarz2017malware}.
\end{packeditemize}

We do not consider the cases that the \cVM intentionally leaks secret, or the security risks brought by unsafe \cVM code, e.g., vulnerabilities in the guest \OS. User applications within the \cVM should implement appropriate security measures for transmitting data across security boundaries, such as encrypting network or disk data by establishing a secure channel through remote attestation.
Since no legacy \OS is written with a malicious hypervisor in mind, the host hypervisor may leverage existing communication interfaces between the host and the guest for attacks. The \TMM could enforce certain security policies similar to Hecate~\cite{ge2022hecate}, but in general we consider the protection of such attacks, e.g., due to hypervisor injected interrupts or exploit of the device interfaces~\cite{li2019exploiting,hetzelt2021via} as orthogonal to our work.
Denial-of-Service (DoS) and side channel attacks~\cite{xu2015controlled, wang2017leaky, lipp2021platypus, li2021cipherleaks, chen2019sgxpectre, zhu2020powerscout, jiang2021quantifying} are also out of scope.

\ignore{
, e.g.,
due to hypervisor injected interrupts or exploit of the device interfaces~\cite{hetzelt2021via}.
the untrusted hypervisor can simply inject an interrupt
while the legacy \OS has interrupt disabled to trigger race conditions
which should have never occurred on a correctly-behaving (virtual)
hardwar
We do not protect the \cVM from  attacks that leverage existing communication interfaces between the host and the guest, e.g.,
due to hypervisor injected interrupts or exploit of the device interfaces~\cite{hetzelt2021via}. }

\section{Virtualizing Arm Confidential Compute Architecture}
\label{sec:design}

\subsection{Overview}

In this section, we provide the overall architecture of \sysname, a virtualized \CCA using TrustZone hardware. To stimulate real world deployment of \sysname on legacy hardware, \sysname is designed as a drop-in replacement of CCA, providing strong compatibility with respect to hardware, firmware and software, strong security guarantees and high efficiency, which are elaborated below. 
\begin{packeditemize}
    \item \textit{Hardware compatibility}. \sysname utilizes features from legacy Arm architecture, and is tailored to work with both the Arm v8.4-A architecture, which includes S-EL2 support, and earlier platforms that lack S-EL2 support. Furthermore, \sysname is compatible with most CCA interfaces at the API level (See Table~\ref{table:tmi}).
    
    \item \textit{Software and firmware compatibility}. Firstly, the \sysname design requires minor modification to the host software infrastructure (i.e., the hypervisor) and the guest VM, without requiring any changes to the application software. As such, \sysname supports running commodity VM images as \cVM with no additional effort. Secondly, due to the strong compatibility with CCA at the API level, the software and firmware stack constructed for \sysname can be reused for CCA with no changes when CCA hardware is available in the future (See Table~\ref{table:software}).
    
    \item \textit{Security}. \sysname aims to provide a similar security guarantee to CCA. \sysname adopts a similar model to CCA to minimize the \TCB, in which the heavy software stack, i.e., the hypervisor runs in the normal world, while the TMM runs as a reference monitor to enforce strong security. In addition, \sysname supports the integrity verification of the \cVM through remote attestation.
    
    \item \textit{Performance}. In contrast to other designs (CCA~\cite{armcca} and Twinvisor~\cite{li2021twinvisor}), the secure memory is managed entirely by the \TMM. \sysname avoids the cost of memory delegation, reducing the complexity in managing the secure memory. 
\end{packeditemize}



\begin{figure*}[htbp]
    \centering
    \begin{subfigure}{0.48\textwidth}
        \centering
        \includegraphics[width=\textwidth]{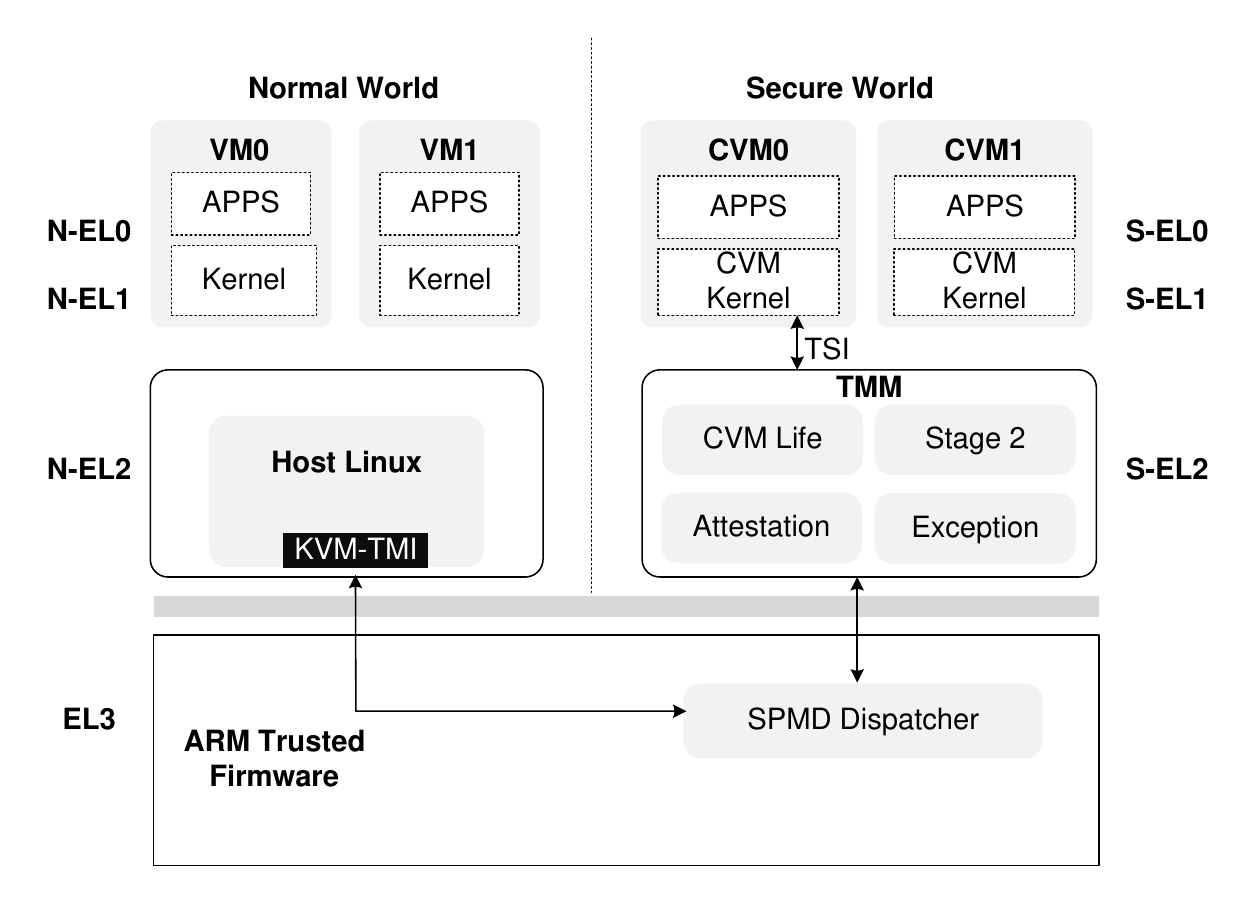}
        \caption{\sysname-SEL2}
        \label{fig:sel-2-arch}
    \end{subfigure}
    \begin{subfigure}{0.4\textwidth}
        \centering
        \includegraphics[width=\textwidth]{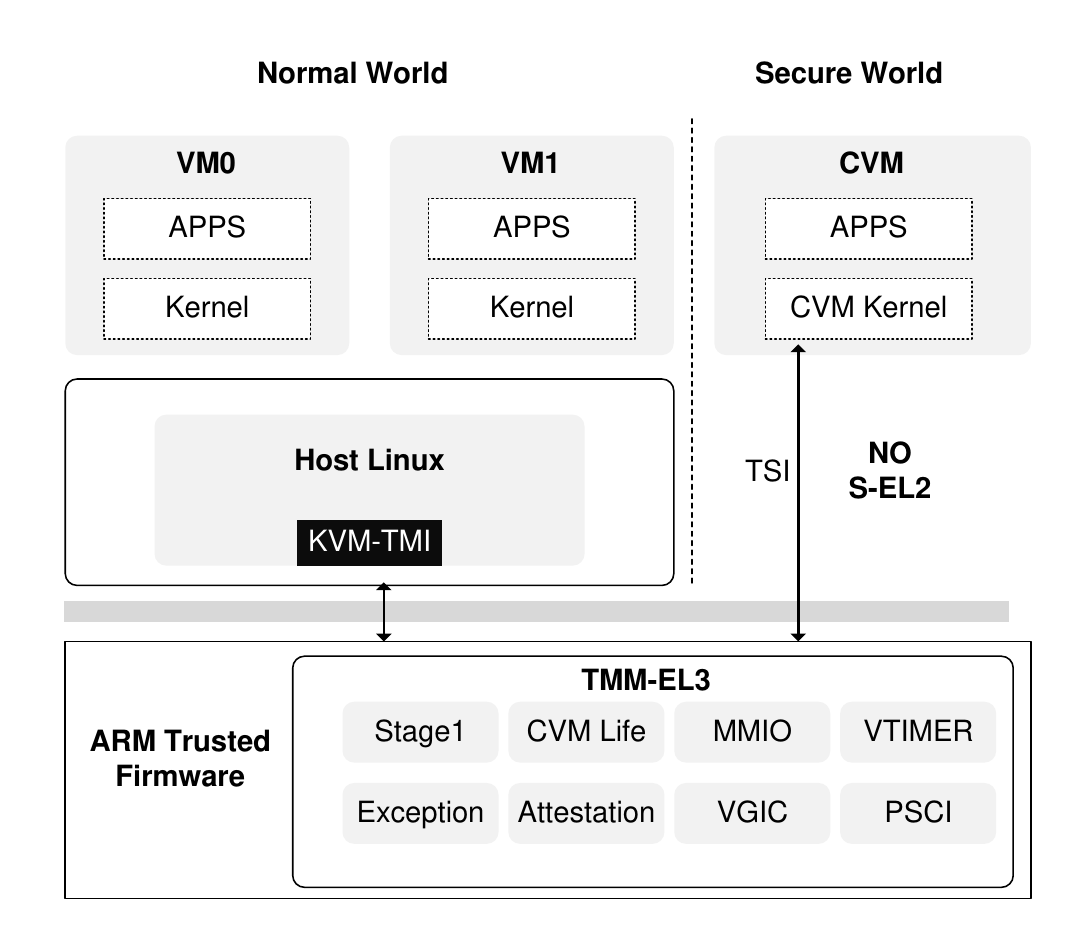}
        \caption{\sysname-EL3}
        \label{fig:s-el3-arch}
    \end{subfigure}
    \caption{The architecture of \sysname.}
    \label{fig:arch}
\end{figure*}

Figure~\ref{fig:arch} illustrates \sysname's architecture. The secure monitor in EL3 enforces memory isolation between the normal and secure worlds, by configuring the \TZASC. The normal VMs and \cvms reside in the normal world and secure world, respectively.
\sysname leaves the complexity of task scheduling and resource allocation to the untrusted hypervisor in the normal world, which manages the life cycle of all normal VMs and \cvms. In particular, a component in the host (named KVM-\sysname) communicates with the secure world for the management of \cvms. 
As the hypervisor lacks direct access to the secure world, a component known as the \TMM serves as a trusted reference monitor, aiding the hypervisor in managing \cvms. Just like in CCA, the \TMM maintains a TrustZone Execution Context (TEC) object, which is associated with a VCPU of the \cVM. The TEC object is used by the \TMM to store the register state of a TEC.


On Arm v8.4-A and later architectures, the \TMM operates at S-EL2 (Figure~\ref{fig:sel-2-arch}); while on former architecture, the \TMM also operates at EL3 (Figure~\ref{fig:s-el3-arch}).
The \TMM enforces memory isolation among all \cvms either by configuring the stage 2 page tables or by write-protecting the \cVM's page table (Sec.~\ref{subsec:isolation}). In Sec.~\ref{subsec:interrupt}, we explain how interrupt and device accesses are emulated. Furthermore, in Sec.~\ref{subsec:attestation}, we discuss the measurement and attestation of \cvms. For brevity, we use \sysname-SEL2 and \sysname-EL3 to represent our systems where the \TMM operates at S-EL2 and EL3, respectively.

The separation of resource allocation and access control necessitates a communication channel between these two worlds. Therefore, \sysname offers a set of TrustZone Management Interfaces (TMIs) to the normal world, and the KVM-\sysname communicates with the \TMM through TMIs. The TMM receives TMI requests from the normal world, manages the state transitions during TMI calls, and returns to the normal world once the TMIs are completed. Furthermore, the TMM offers a collection of TrustZone Service Interfaces (TSIs) for the \cvms, e.g., allowing the \cVM to request the generation of the attestation report. 

\subsection{Memory Isolation}
\label{subsec:isolation}
Memory isolation schemes operate at two levels. Firstly, \sysname utilizes the \TZASC hardware to facilitate the separation of memory between the secure world and normal world. 
Secondly, \sysname enforces memory isolation among mutually distrustful \cvms. In \sysname-SEL2, the \TMM manages the stage-2 page tables of all \cvms, ensuring that the secure memories utilized by the \cvms do not overlap. In \sysname-EL3, the system prevents a \cVM from accessing the memory of another \cVM by designating their page tables as read-only. Consequently, any modifications to the page table made by the \cVM are intercepted and audited by the TMM at EL3.

\para{Cross-world isolation}
In a standard setting, the hypervisor oversees the system's physical memory and utilizes the buddy system to allocate free memory to the VMs. Without specific intervention, the secure memory used by \cvms and the normal memory would be interleaved with each other. However, the \TZASC hardware only allows configuration of security properties for up to 8 consecutive memory regions, making it difficult to support both \cVM and normal VM managed by the host simultaneously. Twinvisor adopts a runtime migration approach for physical memory between secure and normal world. This enables an overcommitment of \cVM memory, thereby enhancing memory utilization. However, due to the limited number of regions supported by \TZASC, Twinvisor has to constantly split and compact memory in order to reduce the total number of continuous secure memory regions. The design inevitably incurs overhead due to the dynamic management of memory mapping and TLB flushing, particularly in scenarios characterized by extensive memory operations.

To address this challenge, the TMM in \sysname takes full responsibility for managing the secure memory used by CVMs. Specifically, \sysname incorporates a secure boot process. Upon system boot, the firmware is the first component to be loaded. A fixed region of memory is reserved for the use as secure memory. The size of this reserved memory can be configured in the BIOS. Subsequently, the firmware designates this memory as reserved, preventing the host from accessing it. Once the firmware completes its initialization, it hands over control to the \TMM.\footnote{The measurement of the firmware, all \sysname components, and the \cvms will be discussed in detail in Sec.~\ref{subsec:attestation}.}

For efficient management of the secure memory, we design the buddy and slab secure memory allocators in \TMM. Buddy allocator can reuse freed memory blocks efficiently, as they can be merged with adjacent free blocks to form larger blocks. Therefore, the buddy allocator is used to reserve a large chunk of memory for a \cVM.
Slab allocator reduces memory fragmentation and improves memory utilization by allocating memory in fixed-size chunks, so it’s used to allocate small piece memory for metadata such as TEC. It is worth mentioning that Non-Uniform Memory Access (NUMA) affinity is also considered in our design, in other words, the memory allocator prioritizes the allocation of memory within the NUMA node
which belongs to physical CPUs assigned to the \cVM. These designs not only effectively overcome the constraints imposed by the \TZASC hardware and allow for the support of an unlimited number of \cvms without requiring dynamic movements and mappings of memory pages, but also improve the system performance by bringing memory nearer to processor cores.

The hypervisor, responsible for the life cycle management of \cvms, retains its role in making decisions of memory allocation and deallocation for \cvms. However, it no longer handles secure memory management. To facilitate collaboration between the hypervisor and the TMM, we introduced a new set of TMIs, in addition to the standard CCA's RMIs. These new TMIs include tmi\_mem\_alloc and tmi\_mem\_free for secure memory allocation and deallocation, and tmi\_map\_ipa\_range and tmi\_unmap\_ipa\_range for mapping and unmapping Intermediate Physical Addresses (iPAs) to System Physical Addresses (sPAs).

\para{Cross-\cVM memory isolation}
On Armv8.4 and subsequent platforms, the isolation of multiple \cvms is achieved by leveraging the stage-2 address translation feature provided by the S-EL2 extension. This enables the \TMM to efficiently manage and safeguard the memory resources allocated to each \cVM, ensuring that they operate within their designated physical memory spaces.
In \sysname-SEL2,
the TMM manages the stage-2 page tables for \cvms, which effectively prevents memory mapping attacks from the host.

In \sysname-EL3, there is no hardware support for stage-2 page mappings in the secure world. 
To support multiple \cvms, our design restricts a \cVM's ability to modify the page table base register by removing relevant instructions (\texttt{msr\_ttbrx\_el1} and \texttt{msr\_tcr\_el1})~\cite{azab2014hypervision,dautenhahn2015nested}. Additionally, we designate the page tables of each \cVM as read-only, preventing the guest OS from mapping the physical memory of other \cvms into its own address space. Any such attempts are intercepted by the TMM, which then verifies the legitimacy of the physical memory. This approach ensures a secure and isolated environment for concurrent \cvms.

Moreover, \sysname-EL3 enforces a one-to-one mapping between the iPA and sPA of \cvms. 
To circumvent the need for extensive modifications to the KVM, we initially establish one-to-one mappings between the Host Virtual Address (hVA) and sPA in QEMU. Subsequently, we transfer the mappings between the iPA and hVA to the KVM. Overall, we introduced approximately 100 lines of code in QEMU for identical mapping while leaving the KVM unchanged. With the one-to-one mapping, the \cVM effectively manages its own physical memory. Since \sysname-EL3 only requires a single level of address mapping, this design eliminates the overhead of memory virtualization.

\subsection{Interrupt and Device Virtualization}
\label{subsec:interrupt}
According to the \RMM specification~\cite{armrmm}, the host should provide a virtual GIC (vGIC) to the \cVM, and is able to inject virtual interrupts using the \GIC \vCPU interface. 


\para{Interrupt translation on \sysname-SEL2}
In \sysname-SEL2, the vGIC is emulated by the host, while the \TMM handles the transitions between different interrupt types of the two worlds.
Specifically, Linux’s Generic \GIC driver assigns the interrupt into Group 1 with Non-secure state. Group 1 interrupts are signaled as either IRQs or FIQs depending on the current security state and exception level. Usually, interrupts are signaled as IRQs when Linux is running at Non-secure EL0/1. However, FIQs are signaled in \cvms while guest OS is executing at \SELzero/\SELone, which are intended to be handled by Linux as IRQs. 

\begin{figure}
\centering
\includegraphics[width=\columnwidth]{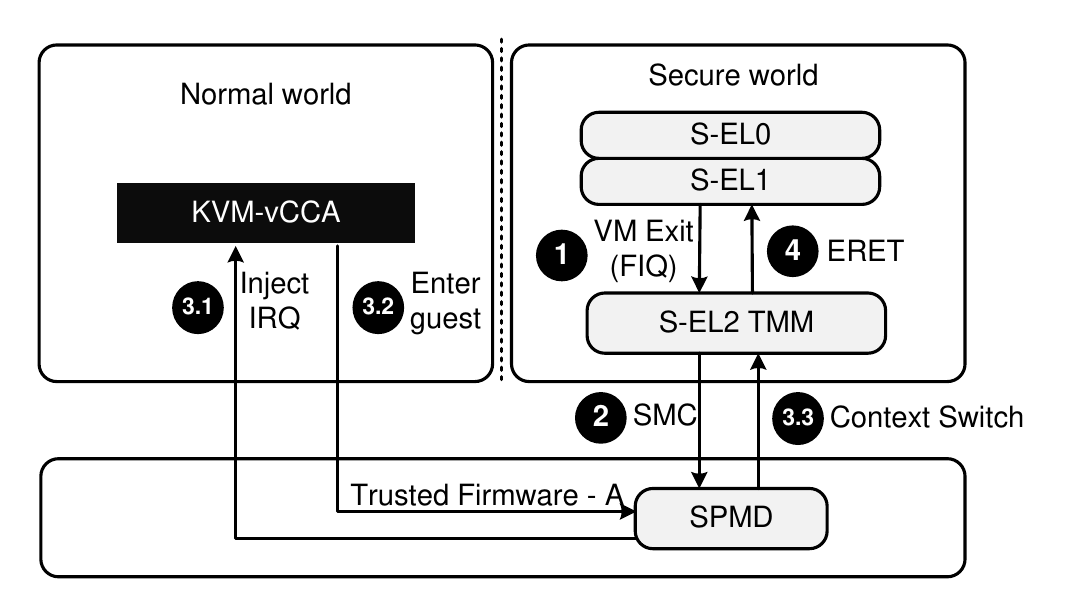}
\caption{\sysname-SEL2 interrupt translation.}
\label{fig:interrupt_sel2}
\end{figure}

A straightforward solution is to modify the interrupt group and security setting by modifying the \cVM's \GIC driver. To minimize the modification of the guest OS,
FIQs captured by \TMM will be bounced back to hypervisor which will handle FIQ as \IRQ and then inject vIRQ to the \cVM. 
The detailed steps are shown in Figure~\ref{fig:interrupt_sel2}: 
\circled{1}while executing at EL1 or EL0, a Group 1 interrupt for the secure Security state is taken as an FIQ, which will be captured by \TMM due to FIQ exit; \circled{2}The \TMM marks the exit reason as \IRQ rather than FIQ, and then acknowledge the KVM to handle physical interrupt as \IRQ via \SMC; \circled{3}The physical interrupt is handled by the KVM which will write one of the list registers to inject vIRQ to the \cVM. \circled{4}Finally, the guest OS handles the incoming vIRQ, when \TMM executes an ERET instruction to enter the \cVM. After the interrupt is handled, the vCPU restores the execution state before the interrupt occurred.

\begin{figure}
\centering
\includegraphics[width=0.9\columnwidth]{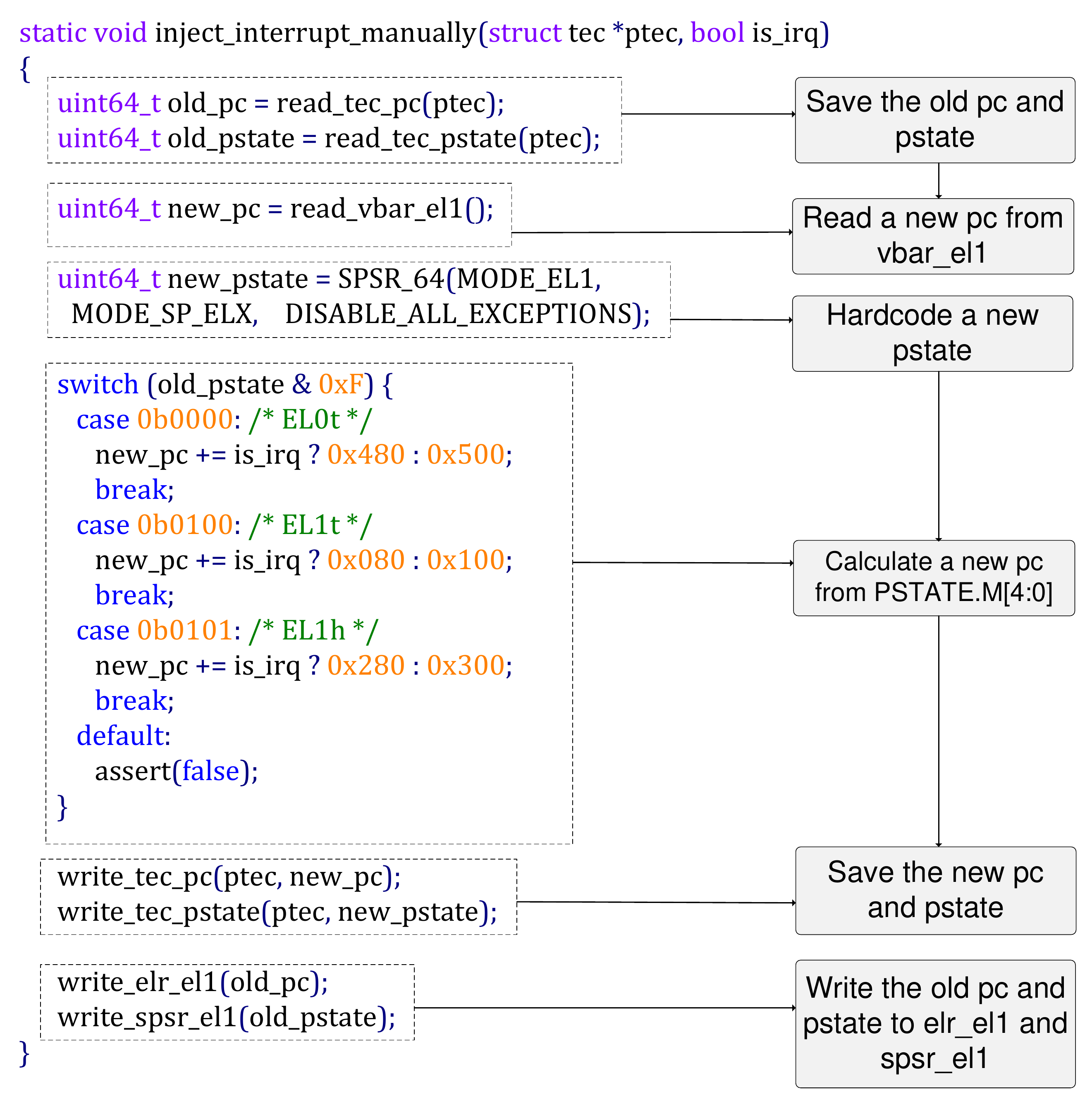}
\caption{\sysname-EL3 software interrupt injection.}
\label{fig:interrupt}
\end{figure}

\para{Interrupt virtualization on \sysname-EL3}
In \sysname-EL3, the interrupt virtualization lacks hardware support in the secure world. Specifically, the \GIC list registers are not supported, which means that the hardware does not track the interrupt state and we cannot inject vIRQ to \cVM by writing these registers. To address these issues, we emulate the registers entirely in software. To maintain the interrupt state, we store the state of the registers in the TEC for each vCPU and model the state transition of the registers based on the type of interrupts. We also emulate the injection of vIRQ to \cVM entirely in the \TMM. As shown in Figure~\ref{fig:interrupt}, we first retrieve the old processor context before the interrupt and construct a new processor state according to the type of interrupts. Then we configure the Exception Link Register (ELR) as the old processor state. Finally we resume the \cVM and the vCPU begins execution based on the new processor state. After the guest OS handles the interrupt, it initiates an ERET instruction, which restores the previous process state from the ELR, marking the end of the interrupt handling procedure. Besides, in cases where the guest OS needs to access list registers, we replace them with SMC calls.

\para{MMIO virtualization}
In the CCA model, peripheral devices can be assigned to a \cVM, and only the assigned \cVM has access to the device interface. The \cVM may access the assigned peripheral devices through the Memory-mapped Input/Output (MMIO) interface.
On \sysname-SEL2, MMIO accesses triggers translation faults which are captured by the \TMM running at S-EL2. The hardware will set the Exception Syndrome Register (ESR) register to encapsulate the MMIO access information, and the MMIO fault is handled by the hypervisor. On \sysname-EL3, we modify the guest OS by replacing MMIO accesses with SMC instructions that will trap to the \TMM. Without hardware support, the \TMM updates the ESR as per the Arm specification.
Similar to virtualizing the MMIO interface, the \TMM is also responsible for the emulation of Power State Coordination Interface (PSCI) and timer etc. The details are omitted here.



\subsection{Measurement and Attestation}
\label{subsec:attestation}


\para{Root of trust}
Our design relies on an on-chip Root of Trust (\RoT), such as a Trusted Platform Module (TPM). This privileged hardware module is initialized immediately after the SoC reset is released and offers runtime APIs. The \RoT serves three primary functions: \RoT for measurement, \RoT for storage, and \RoT for reporting.
The \RoT for measurement refers to the immutable code of the Boot Loader stage 1 (i.e., BL1) in the ATF.
The BL1 code is pre-installed in the system's read-only memory (ROM) and is executed at the beginning of the boot sequence to initiate the measurement process. This ensures that the \RoT for measurement is activated, even if the system has been compromised.
During the boot process, intermediate measurements are securely stored in the \RoT for storage. This storage area can only be written by privileged code using the extend operation.

The \RoT for reporting is responsible for signing the measurement. It utilizes the private key of the platform's asymmetric key pair, known as the Initial Attestation Key (IAK), which is securely stored within the RoT's secure key storage.
The IAK, along with its certificate, is securely provisioned through a certificate authority (CA) and pre-installed during the platform's fabrication process. This unique key pair is used to cryptographically bind the platform measurement report (referred to as the report) with the platform's identity.

\para{Measured boot and token generation}
The measurement process encompasses both the platform and \cVM measurements, complying with CCA specifications.
During system boot, the BL1 code employs a hashing algorithm to measure the BL2 code of ATF before executing it. This process also involves measuring platform configurations, such as the life cycle state and hardware setup.
Following that, the BL2 code measures each component of the BL3 codes, including the TMM. Before executing the TMM, the \cVM attestation key (CAK) is generated to delegate the signing of \cVM tokens to the TMM.
The platform measurements mentioned above, along with the platform identity and lifecycle state, form the platform token, which is then signed by the IAK.

\begin{table}
\caption{Compatibility of TMM interfaces.} 
\label{table:tmi}
\centering
\small{
\begin{tabular}{C{1.9cm} C{2.7cm} C{0.8cm} C{1.7cm}} \toprule
    \textbf{Interface types} & \textbf{Description} & \textbf{Count} & \textbf{Compatible?} \\ \midrule
    TrustZone Management Interface & Used by the hypervisor to manage \cvms & 19 & Yes \\ \midrule
    TrustZone Service Interface & Used by the \cvms software to request services from the \TMM & 10 & Yes \\ \midrule
    Power State Control Interface & Trusted handling of \cVM's PSCI events & 8 & Yes \\ \midrule
    TrustZone Management Interface & (\textit{New}) Used by the hypervisor to request and mapping \cVM's memory & 4 & No \\
     \bottomrule
\end{tabular}}
\end{table}


During the creation of a \cVM, the TMM meticulously records each configuration step, leading up to the \cVM's creation. This includes measuring and cryptographically extending the initial code and data of the \cVM to form the \cVM's initial measurement (CIM). Specifically, the creation of a \cVM initializes the CIM, while the loading of a data page and a TEC extends the CIM.
Once the \cVM is activated, the CIM is protected by the TMM and remains immutable until the \cVM is destroyed.
Additionally, an interface named TSI\_MEASURE\_EXTEND is provided to the guest OS, allowing the OS to measure the applications and securely store the results in the TMM's extensible measurement slots.
The \cVM token encapsulates both the initial and extended measurements of the \cVM, and it is signed using the CAK.

\begin{table}
\begin{threeparttable}[b]
\caption{Compatibility of software stack.} 
\label{table:software}
\centering
\small{
\begin{tabular}{C{1.7cm} C{1.5cm} C{2.8cm} C{1.3cm}} \toprule
    \textbf{Components} & \textbf{Belonging to} & \textbf{Description} & \textbf{Reusable?} \\ \midrule
QEMU & Hypervisor & Virtual Machine Manager for KVM & Yes \\ \midrule
KVM & Hypervisor & \cVM life cycle management & Yes\tnote{$\dagger$} \\ \midrule
Exception handling & Hypervisor & Handling exceptions triggered during \cVM execution & Yes \\ \midrule
\cVM-TMM interface & \cVM & \cVM Software stack for requesting TMM services, such as remote attestation & Yes \\ \midrule
Secure memory management & \TMM & Management of \cvms's secure memory & Partially\tnote{$\ddagger$} \\ \midrule
Interrupt and device virtualization & \cVM and \TMM & See Sec.~\ref{subsec:interrupt} & No \\
\bottomrule
\end{tabular}}
\begin{tablenotes}
\item[$\dagger$] The hypervisor needs to requests secure memory for a \cVM through the new TMIs, but most of the hypervisor code is reusable.
\item[$\ddagger$] The components for managing stage-2 page tables and handling stage-2 page faults are reusable, while the buddy system and slab based secure memory allocator (Sec.~\ref{subsec:isolation}) are not.
\end{tablenotes}
\end{threeparttable}
\end{table}

The platform token and \cVM token are securely bound together by incorporating a hash of the CAK's public key into the platform measurement conducted in the previous stage. This combined data is subsequently signed by the IAK, establishing a immutable linkage between them.

\para{Attestation}
When a remote user (i.e., the verifier) intends to establish a secure connection with the \cVM, they send a challenge value to the \cVM. The \cVM incorporates this challenge value into its measurement report to ensure freshness and returns the signed report. Additionally, the certificate of the AIK is also included in the response.
Upon receiving the report, the verifier first checks the validity of the certificate and then validates the integrity of the report.
Then the verifier retrieves the reference values and compares the measurement against these reference values. It is worth noting that both the ATF and TMM do not have access to the private key of the AIK. 
Even if these components are compromised, attackers cannot create fake reports that would pass verification.


The initial \cVM image, whose integrity can be assured through remote attestation, is considered to not contain any secret information.
During remote attestation, an authenticated key exchange protocol is utilized to establish a shared secret key. This key is used to secure subsequent communications, enabling users to securely provide confidential data to the \cVM for processing.


\subsection{Software Compatibility Analysis}
In this section, we provide a summary of \sysname's compatibility with CCA software stacks.

\para{TMM interfaces}
As shown in Table~\ref{table:tmi}, most TMM interfaces in \sysname, including TMIs and TSIs, are designed to be compatible with CCA's specifications~\cite{armrmm}. Furthermore, \sysname introduces four new TMIs for secure memory allocation and deallocation to \cvms, as explained in Sec.~\ref{subsec:isolation}.

\para{Software stack}
As shown in Table~\ref{table:software}, the majority of CCA software stacks, such as the hypervisor, \cVM, and RMM, can be reused in \sysname. However, \sysname introduces several new components, including the \TMM's management of secure memory, as well as interrupt and device virtualization.

\section{Security Analysis}
\label{sec:security}

\begin{table}
\caption{LoCs of \sysname's software stack.} 
\label{table:loc}
\centering
\small{
\begin{tabular}{C{2.7cm} p{3.8cm} C{1cm}} \toprule
    \textbf{Component} & \multicolumn{1}{c}{\textbf{Description}} & \textbf{Line of Code} \\ \midrule
    \sysname-SEL2 \TMM & Manage the execution enviroment of \cvms with buddy and slab memory allocator & 4.0K \\
    \sysname-EL3 \TMM & Manage the execution enviroment of \cvms with buddy and slab memory allocator & 3.2K \\
     \sysname-SEL2 \cVM & Modified Linux \cVM guest kernel with SWIOTLB bounce buffer and remote attestation & 1.3K \\
     \sysname-EL3 \cVM & Modified Linux \cVM guest kernel with SWIOTLB bounce buffer, remote attestation, MMIO with SMC and interrupt handle with SMC & 2.1K \\
    KVM-\sysname & Modified Linux KVM communicating with \TMM & 1.8K \\
    QEMU & Arm64 emulator and KVM VMM supporting \sysname & 2.4K \\
    TEST & Bare Metal testing framework to test \TMM interfaces & 5.3K \\ \midrule
    \normalsize
    \textbf{Total} &  & \textbf{20.1K} \\ \bottomrule
\end{tabular}}
\end{table}

\para{\TCB} Our implementation is built upon the TMM Dispatcher(TMMD) and SPMC framework.
In total we made 20.1K lines of code (LoCs) insertions or changes, as shown in Table~\ref{table:loc}. We stress that most of our code base can be reused by CCA with only a few changes with the help of pre-defined macros. For example, \TMM uses Virtualization Secure Translation Table Base Register (\path{VSTTBR_EL2}) as the base register for stage 2 of the S-EL1\&0 translation regime, while CCA uses the Virtualization Translation Table Base Register (\path{VTTBR}) register as the base register.

The \TCB includes the TrustZone hardware, the \ATF and the \TMM. In total, the TMM we added to the ATF is only about 4.0K LoCs for \sysname-SEL2, and 3.2K LoCs for \sysname-EL3, which is within the range suitable for formal verification.


\para{Untrusted host}
The host is not allowed to access the \cVM, since the non-secure memory and secure memory are isolated by configuring the \TZASC. 
The \cVM exit events (such as interrupt and exceptions) are first trapped to the \TMM. The \TMM decides whether to pass control to the host, after the \cVM states are safely saved and sensitive information is cleared.
\sysname is designed to provide the same level of protection as CCA upon interrupts and exceptions. That is, the conditions to pass control and the values passed to the host are exactly the same for \sysname and CCA. Similar to existing VM-based TEEs, \sysname does not protect from the attacks that the host hypervisor injects spurious interrupts, which our threat model explicitly permits (see Sec.~\ref{subsec:threatmodel}).

\para{Untrusted \cVM}
The memory isolation among \cvms is ensured by configuring the stage-2 page tables of all \cvms. On platforms without S-EL2, the stage-1 page table of the \cVM is managed by the TMM to prevent a malicious \cVM from accessing physical memory other than its own (Sec.~\ref{subsec:isolation}). Further, we ensure that all \cVM memory are cleared when the \cVM is destroyed to prevent residue data leakage.

\para{\DMA attacks}
The DMA controller is used to allow I/O devices to directly access physical memory. An attacker may try to access the secure memory leveraging \DMA. TrustZone protects against DMA attacks, such that the DMA controller handles secure and non-secure events simultaneously, and prevents non-secure DMA access of secure memory. Built upon TrustZone, \sysname is resistant to DMA attacks. 



\section{Evaluations}
\label{sec:eva}


\begin{table}
\caption{\sysname prototyped system’s configurations.} 
\label{table:configuration}
\centering
\small{
    \begin{tabular}{C{1.8cm} C{3.6cm} C{1.7cm}} \toprule
        \textbf{Component} & \textbf{HOST} & \textbf{\cVM} \\ \midrule
        OS & Ubuntu 22.04 & Ubuntu 20.04 \\
        Kernel & 5.10 & 5.10 \\
        QEMU & 6.2.0 & - \\
        TF-A & 2.7 & - \\
\bottomrule
    \end{tabular}}
\end{table}

\para{Experimental setup}
We evaluated \sysname on real Arm servers. Specifically, for \sysname-SEL2, we used an early version of a real Arm V8.4 server 
with 64 cores running at 2400 MHZ, which is equipped with S-EL2 support and 64 GB RAM. For \sysname-EL3, we used an Arm server with Kunpeng 920 CPU and 256 GB RAM without S-EL2 support. In \sysname-EL3, the CVM's page tables are marked as read-only. 
Table~\ref{table:configuration} shows the software stack of the host.
We modified the Linux KVM hypervisor to work with \TMM, and the changes are quite modest. We decoupled vanilla KVM and \cVM KVM in our implementation, which supports normal VMs and confidential VMs on the same platform. Each \cVM was allocated 2 {vCPU}s and 512\MB RAM, and ran Ubuntu 20.04 with kernel version 5.10. Arm VHE mode was enabled for all measurements to reduce the virtualization overhead. 



\ignore{The evaluations were conducted to answer the following questions:
\begin{packeditemize}
    \item Do TMM interfaces conform to the CCA specifications? (see Sec.~\ref{subsec:test})
    \item How much is the performance gain of using direct mapping, compared with using dynamic mappings adopted by CCA and Twinvisor? (see Sec.~\ref{subsec:evalmappings})
    \item How much overhead is introduced by \sysname to the \cVM, compared with running normal guest VM in the host? (see Sec.~\ref{subsec:evalhost})
\end{packeditemize}}

\ignore{
Although ARMv8.4 specification has been released for years, there are no commercially available ARMv8.4 (with S-EL2 extension) hardware at the time of writing.
In this paper, we run the evaluating experiments in QEMU. As shown in Table~\ref{table:configuration}, the host machine running QEMU is equipped with Intel i9-10900K CPU and 32\GB RAM. We have run both micro benchmark and application benchmark on vanilla KVM and \sysname KVM.
QEMU, acting as the emulator for ARMv8.4 (S-EL2) hardware, is running with max type CPU, 4 VCPUs and 3054\MB RAM.
In the meantime, another QEMU which is running in host Linux acts as VMM to launch the guest VM. It runs with 2 VCPUs and 512\MB RAM. To reduce the overhead of virtualization, Arm VHE mode was enabled for all measurements.
}

\subsection{Sanity Check of TMM Interfaces}
\label{subsec:test}

We implemented the \test framework to ensure that \sysname is consistent with the CCA specifications.
The \test framework runs in bare metal EL2 in the normal world and invokes TMIs to interact with the TMM.
There are 18 test cases covering all TMIs.
The test cases cover multi-core handling, race condition, input sanity checks, various levels of \cVM page tables, inter-world memory sharing and isolation, inter-\cVM isolation, and timer/interrupt handling. The \test framework can be used to test both \sysname and CCA implementations as their interfaces are compatible.

\begin{table}
\caption{Descriptions of application benchmarks. We used the default parameters except those specified in the description.} 
\label{fig:application_benchmark}
\centering
\small{
    \begin{tabularx}{0.5\textwidth}{p{1.25cm} p{6.9cm}} \toprule
        \textbf{Item} & \textbf{Description} \\ \midrule
        MongoDB & MongoDB (v3.6.8) is a document-oriented database program, handling requests from a remote client (YCSB v0.17.0 workload A) with default parameters (operation count: 500000, threads: 16). \\
        Redis & Redis (v5.0.7) is Remote Dictionary Server, handling requests from a remote redis benchmark client (redis-tools) with parameters (requests: 100000, parallel connections: 50, and pipeline requests: 12). \\
        Memcached & Memcached (v1.5.22) is a distributed memory-caching system, handling requests from a remote memtier (v1.4.0) client with parameters (threads=16). \\
        HackBench & Hackbench is both a benchmark and a stress test for the Linux kernel scheduler, running with parameters (loops: 500 and groups: 20). \\
        Apache &  Apache HTTP Server (v2.4.41) is a web server, handling requests from a remote ApacheBench (v2.3) client with parameters (concurrency: 1000 and requests: 500000). \\ 
         \bottomrule
    \end{tabularx}}
\end{table}

\ignore{
\para{Micro-benchmarks using memory operations between the host and \cVM}
In this evaluation, we measured the performance of memory copy from the host to the \cVM, with static mapping and dynamic mapping respectively. The results show that static memory is 12\% to 178\% faster depending on buffer sizes (see Table~\ref{table:memcpy}).
The slowdown is caused by memory address access control checks, mapping and unmapping of host page into \cVM page table.
This is especially significant when the buffer size is small.
For example, during one of the most frequently used TMIs, \path{tmi_tec_enter}, a small amount of context information is passed between the host and TMM.
We note that this benchmark does not measure the impact of TLB flushes induced on \textit{subsequent executions} after the memory copy.
Having that considered, the actual slow down induced by dynamic mapping can be even worse.

\para{Real-world applications}
We evaluated static mapping and dynamic mapping with 3 real-world I/O-intensive applications, i.e., FileIO, MongoDB and Redis. The details for the evaluations are summarized in Table~\ref{fig:application_benchmark}. As shown in Figure~\ref{fig:mapping_macro}, static mapping brings 8\%-25\% improvement over dynamic mapping.
}

\subsection{Micro-benchmarks}

\ignore{
\begin{table}
\caption{Descriptions of micro-benchmarks.} 
\label{fig:micro_benchmark_desc}
\centering
\small{
    \begin{tabularx}{0.48\textwidth}{p{0.5cm} p{7.3cm}} \toprule
        \textbf{Item} & \textbf{Description} \\ \midrule
        \IPI & Signal a Linux kernel ``Function call interrupt'' (IPI1) from VCPU0 to VCPU1. Measure the time interval between VCPU0 fires the interrupt and VCPU1 handles the empty callback function. \\
        I/O & Read MAC address from the virtio-net frontend in the \cVM, and measure the time interval. \\ \bottomrule
    \end{tabularx}}
\end{table}
}

\begin{table}[t!]
\caption{A comparison of basic context switches between vanilla KVM and \sysname KVM.} 
\label{fig:micro_benchmark}
\centering
\footnotesize{
    \begin{tabular}
    {C{1.1cm} C{1.4cm} C{1.4cm} C{1.3cm} C{1.4cm}} \toprule
        \textbf{Operations} & \textbf{vanilla SEL2 KVM} & \textbf{\sysname-SEL2 KVM} & \textbf{vanilla EL3 KVM} & \textbf{\sysname-EL3 KVM}\\ \hline 
        \IPI & 1.93 us &  3.42 us& 2.65 us & 7.26 us \\
        I/O & 18.72 us & 21.34 us & 22.08 us & 39.285 us\\
        \bottomrule
    \end{tabular}}
\end{table}

We designed the micro-benchmarks to evaluate the cost of frequently-used hypervisor operations, including virtual inter-processor interrupt (IPI) and I/O request 
for vanilla KVM and both \sysname-SEL2 and \sysname-EL3 KVM.
We collected the average results over 1,000,000 measurements (see Table~\ref{fig:micro_benchmark}). Vanilla SEL2 and vanilla EL3 KVM represent running VMs in the normal world on the corresponding \sysname-SEL2 and \sysname-EL3 platforms.  


For virtual \IPI, we issued an \IPI from {\vCPU}0 to {\vCPU}1, which triggered a remote callback function on {\vCPU}1. {\vCPU}0 thread waited until the function returns and counted the elapsed time. The entire process involved two round trips of interrupt emulation, as the physical \IPI will cause another VM exit from target {\vCPU}1 to the \TMM, while vanilla KVM only involved transitions between the guest VMs at EL-1 and the hypervisor at EL-2. As such, we observed that \IPI for \cVM took more than twice of that for vanilla KVM. The additional cost comes from the world switches.

In the I/O benchmark, we measured the time required to get the MAC address from the \cVM through the virtio-net interface, which gets the response from the QEMU virtio-net device model in the host. The entire process includes the similar HVC cost (i.e., the round trip from \cVM to KVM), and additional \vring synchronization cost in the \texttt{\virtio} framework.


\ignore{\begin{figure*}
\centering
\includegraphics[width=0.85\textwidth]{figure/fio-6.png}
\caption{FIO test with random/sequential read and write operations.}
\label{fig:fileop}
\end{figure*}
}

\begin{figure}[!ht]
\centering
\includegraphics[width=0.42\textwidth]{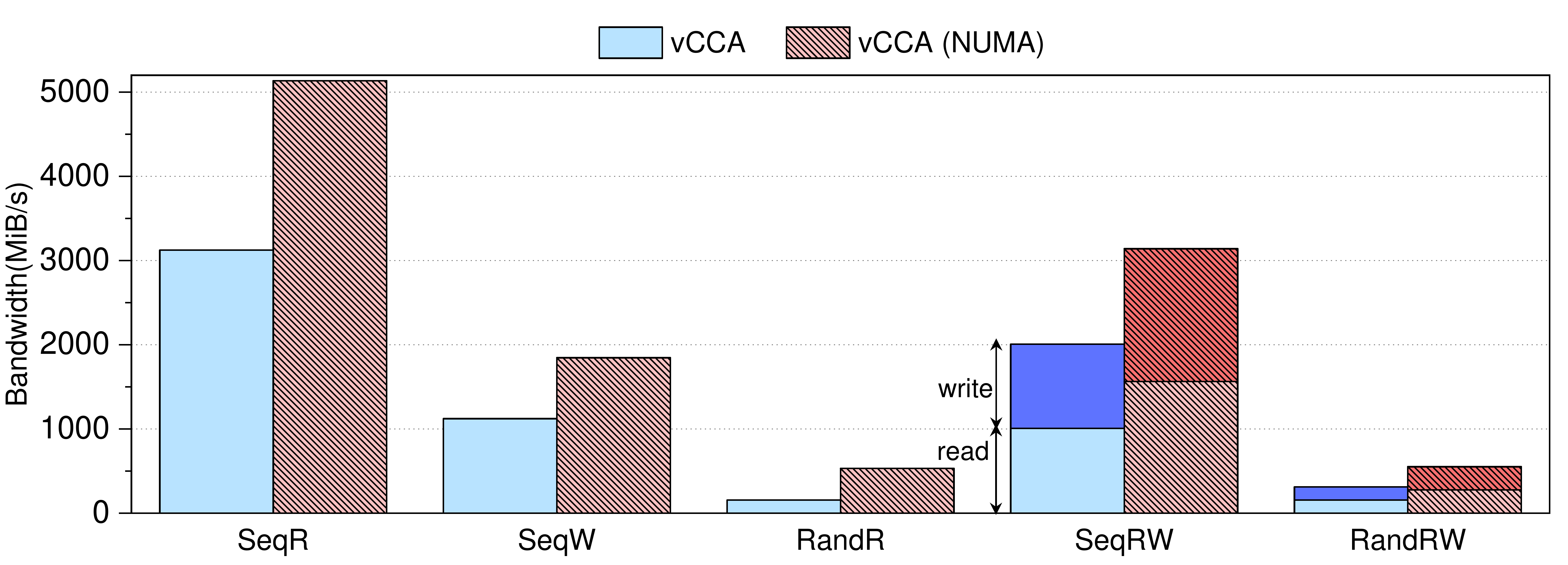}
\caption{FIO test with random/sequential read and write operations.}
\label{fig:fileop}
\end{figure}

\para{Secure memory management with NUMA affinity}
To evaluate the effectiveness of NUMA affinity, we evaluated the I/O performance of \sysname using FIO (v3.33) in its default configuration. We conducted randomized and sequential read and write operations on the disk using a single thread, and measured the memory bandwidth.
We selected the libaio engine, which utilizes the Linux native asynchronous I/O interface, and set the IO depth to 32.
Each test involved accessing a 10 GB file, with a block size of 4 KB for random read and write operations, and a block size of 1 MB for sequential read and write operations. For the evaluations involving both read and write operations, a 1:1 ratio was set.
The test duration was set to 60 seconds, with a ramp time of 15 seconds to ensure stable results.
As shown in Figure~\ref{fig:fileop}, 
with the NUMA optimization, the performance improved by approximately 66.7\%.

\ignore{
We issued hypercall from \cVM to read Arm\underline{~}SMCCC\underline{~}VERSION\underline{~}FUNC\underline{~}ID in the hypercall micro-benchmark.
Hypercall in vanilla KVM only involves \EL trap cost, however, it needs additional steps to finish a hypercall in \sysname. To analyze the overhead of hypercall for \sysname, we break down the time cost for each step: 
1) trap from \cVM in \SELone to \TMM in S-EL2 (4 us); 
2) trap from \TMM to \ATF for world switch (27 us);
3) save \cVM context and restore NS context (30 us); 
4) ERET to KVM in EL2 (4 us); 
5) KVM calls \TMI interface to trap to \ATF again (23 us); 
6) save NS context and restore \cVM context (30 us); 
7) ERET to \TMM in S-EL2 (4 us);
8) copy data from TEC enter’s general purpose registers (30 us);
9) ERET to \cVM in \SELone (4 us). From the analyze, we could see the roundtrip transitions (steps 2, 3, 4, 5, 6, 7, 8) between \TMM and KVM takes approximately 148 \us, however, these steps are not involved in vanilla KVM. The remaining 94 us comes from other operations, such as timer state and error checking.

In the virtual \IPI micro-benchmark, an \IPI is sent from VCPU0 to VCPU1, which will invoke a remote callback function on VCPU1 and wait until the function returns. 
Similar to the roundtrip transitions for one single interrupt emulation as shown in Figure~\ref{fig:interrupt}, \IPI sending involves two roundtrips, because the physical \IPI will causes another VM exit from target VCPU1 to \TMM. We could see \IPI sending for \cVM will take more than twice times of that for vanilla KVM, and the cost is coming from the transitions of world switch. However, vanilla KVM only involves transitions between the guest VMs at EL-1 and the hypervisor at EL-2.

In the I/O micro-benchmark, we read the MAC address from the virtio-net frontend driver in \cVM, which will get MAC information in the QEMU virtio-net device model in the host. The I/O process includes the same steps from \cVM to KVM and back as hypercall, the different is that it involves  \virtio's \vring synchronization cost. However, it will not dominate the I/O overhead, since \vring is not synced every time for each \cVM enter and exit due to the  \virtio’s on-demanding working style.
}

\begin{figure*}[h]
\center
\begin{subfigure}[b]{0.24\textwidth}
\centering
\includegraphics[width=\linewidth]{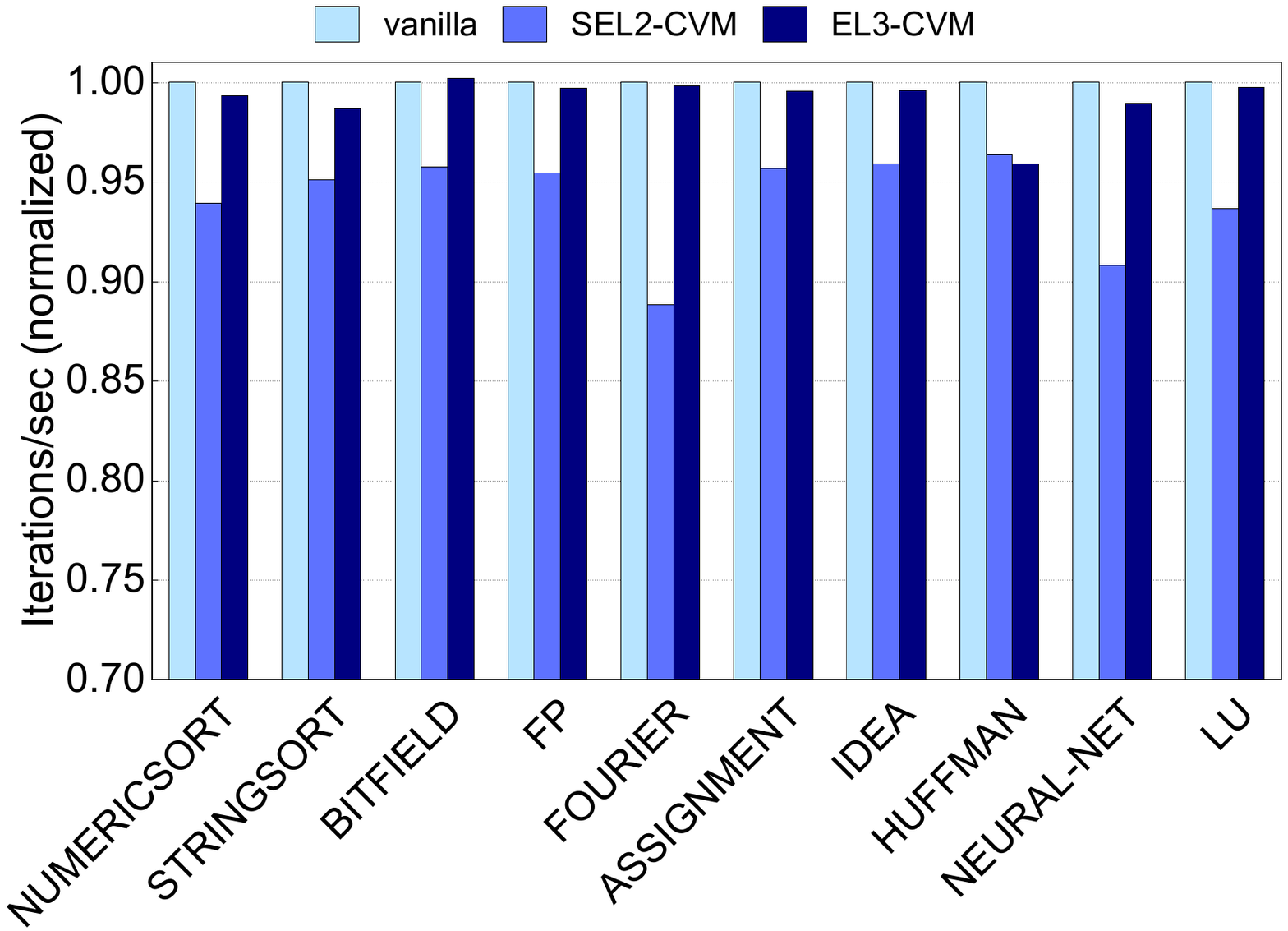}
\caption{Linux/Unix nbench}
\label{fig:nbench}
\end{subfigure}
\begin{subfigure}[b]{0.24\textwidth}
\centering
\includegraphics[width=\linewidth]{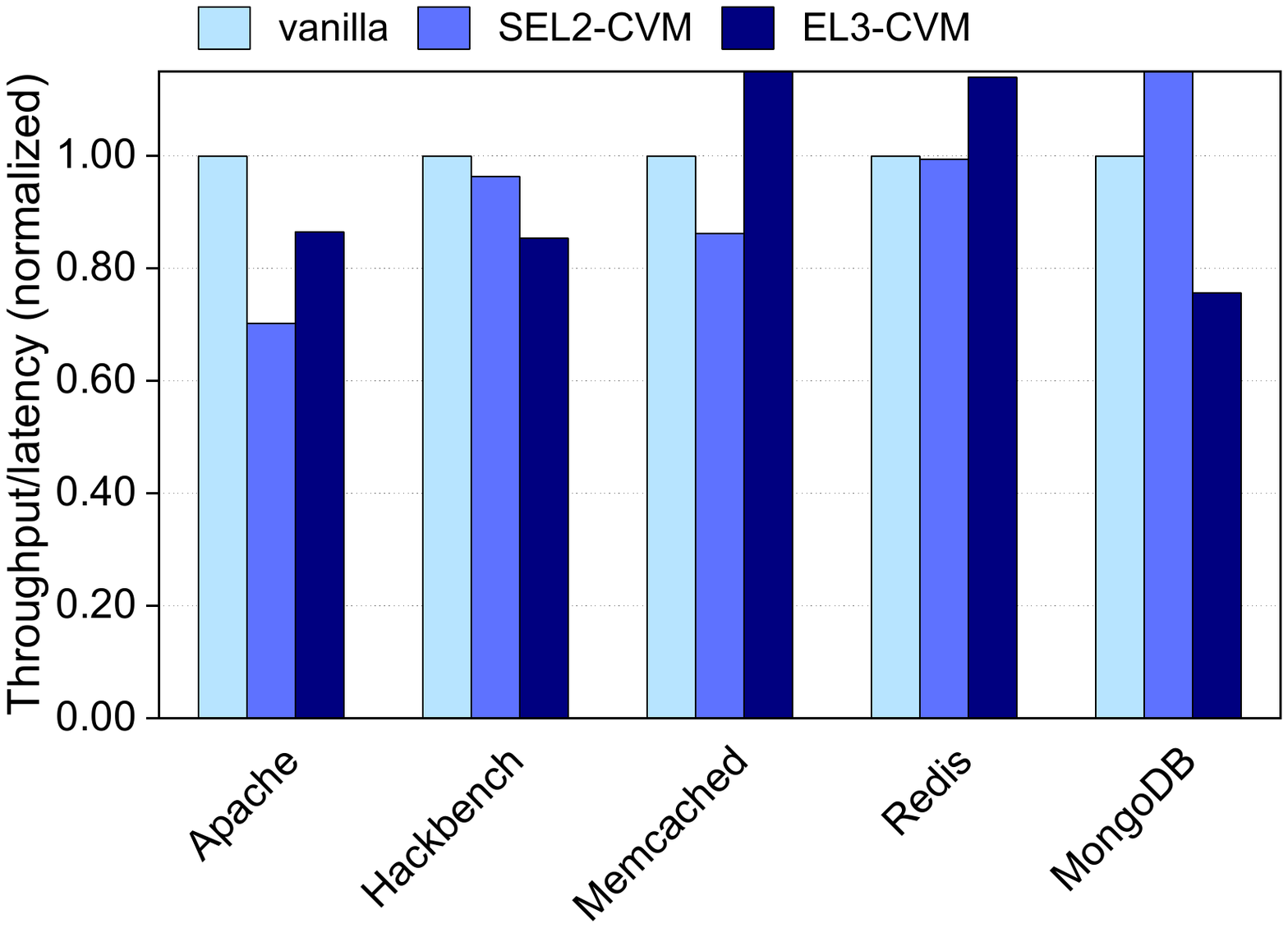}
\caption{Application benchmarks}
\label{fig:appbench}
\end{subfigure}
\begin{subfigure}[b]{0.24\linewidth}
\centering
\includegraphics[width=\linewidth]{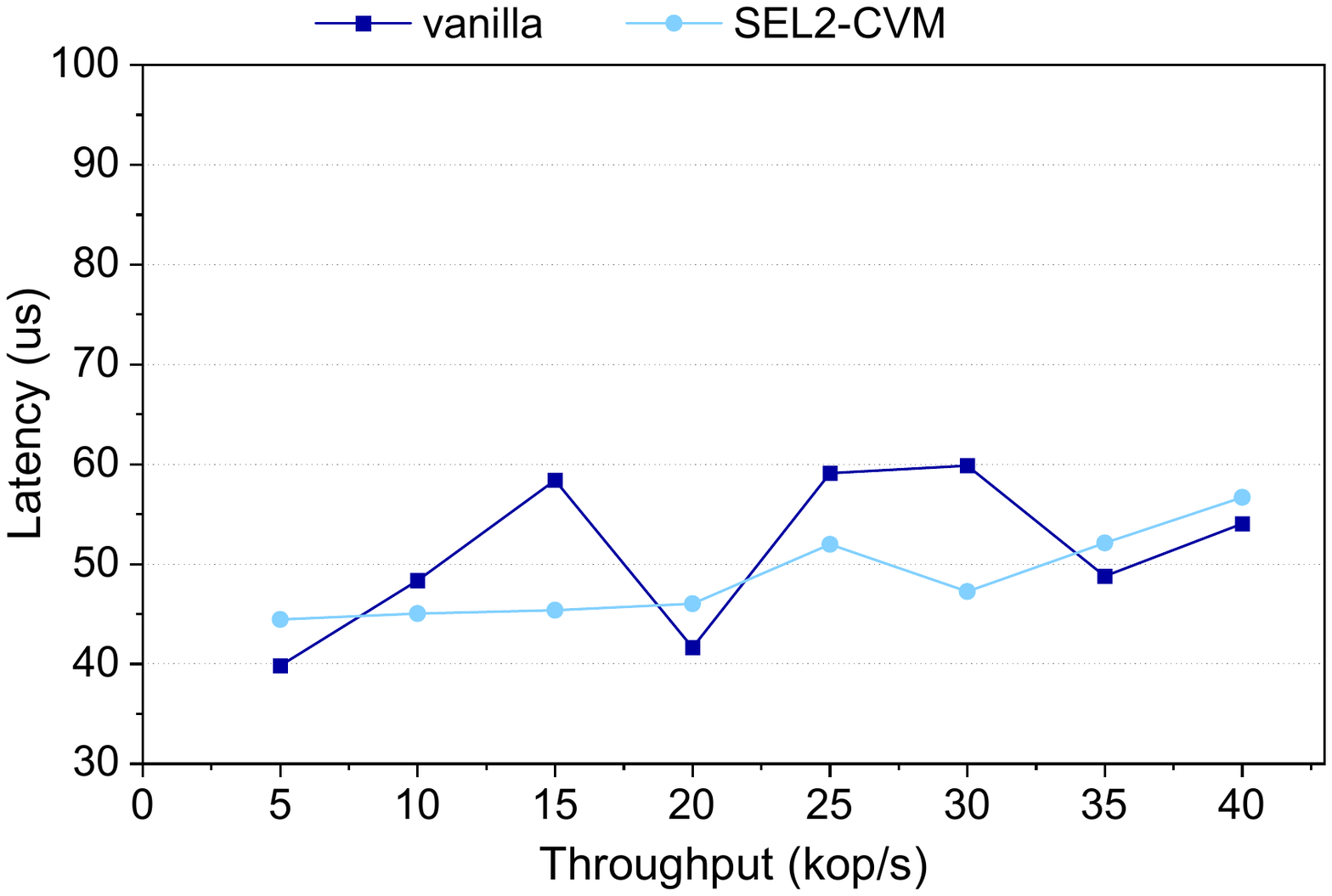}
\caption{Redis-SEL2}
\label{fig:redis-sel2}
\end{subfigure}
\begin{subfigure}[b]{0.24\linewidth}
\centering
\includegraphics[width=\linewidth]{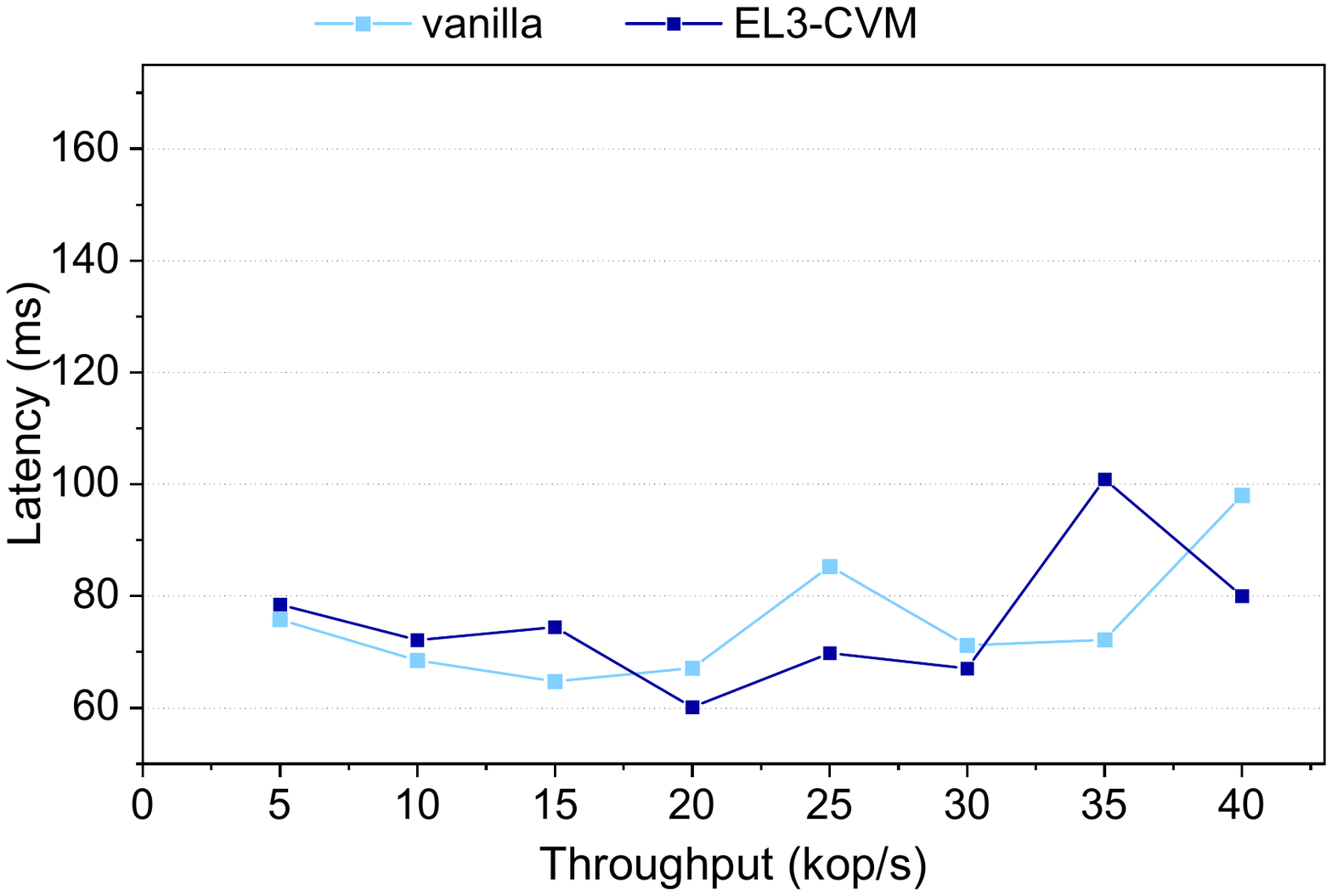}
\caption{Redis-EL3}
\label{fig:redis-el3}
\end{subfigure}
\caption{Macro-benchmarks between \sysname and vanilla KVM. The absolute values for application benchmarks in the format [vanilla-SEL2, \sysname-SEL2, vanilla-SEL2-EL3, \sysname-EL3] are listed as follows: \circled{1}Apache requests/s [18703.14, 15885.28,10534.53,13236.54]; \circled{2}Hackbench in total times (s) [0.5638,0.5848,0.7811,0.9147]; \circled{3}Memcached in Ops/s [493313.64,425533.23,267982.9,450717.4]; \circled{4}Redis in SET requests/s [1157935.94,1150683.95,683599.5,784141.9]; \circled{5}MongoDB in throughput (Ops/s) [1930.33,2257.75,1195.03,903.72].}
\label{fig:eval_macro}
\end{figure*}

\subsection{Macro-benchmarks}
\label{subsec:evalhost}

\para{Linux/Unix nbench}
Linux/Unix nbench is a benchmark suite for the evaluation of a computer system's CPU, FPU, and memory system. It specifically excludes I/O operations and system calls related measurements. Here, we utilized nbench to assess the performance of the CPU, FPU, and memory system within the \cVM. The results in Figure~\ref{fig:nbench} indicate that \sysname introduces an average overhead of approximately 1.8\% for \sysname-SEL2 and 0.8\% for \sysname-EL3.

\para{Real-world Applications}
We performed the application benchmark as described in Table~\ref{fig:application_benchmark}. The Y-axis of Figure~\ref{fig:appbench} represents performance normalization relative to running in the vanilla KVM (the higher the better). We conducted real-world evaluations by running each test five times and collecting the averaged values. The overhead observed in real-world applications is notably lower compared to that of micro-benchmarks. This difference arises because state transitions make up only a small portion of the overall execution time.

In \sysname-SEL2, the overhead for Hackbench remains below 10\% compared to the vanilla KVM. For I/O-intensive workloads like Apache, the average overhead reaches approximately 29.7\%. These workloads necessitate frequent MMIO and interrupt emulations, leading to a higher number of \cVM exits and world switches. Furthermore, they involve frequent interactions with external devices, such as networking, which are implemented using the \virtio framework. 


In \sysname-EL3, the overhead for Memcached is about 33.8\% compared to the vanilla KVM. \sysname outperforms the baseline in other application benchmarks. For example, Hackbench and Redis exhibit improvements of 64.3\% and 16.3\%, respectively, compared to the baseline. This can be attributed to the fact that in \sysname-EL3, \cvms operate in bare-metal mode, eliminating the overhead of CPU and memory virtualization. Although the CVM's page tables are marked as read-only, we confirmed that the number of page table operations is small during the evaluation and does not cause significant overhead.


To further evaluate performance in scenarios involving both memory and I/O intensiveness, we conducted a Redis latency test with varying throughput. We loaded 10 MB of data using YCSB workload A and executed 10,000 operations from 10 concurrent clients over the local loopback. We progressively increased the request frequency and measured the corresponding latency at different throughput.
Figure~\ref{fig:redis-sel2} and Figure~\ref{fig:redis-el3} illustrate the observed latency at different achieved throughput by increasing the request frequency. The results indicate that, in most cases, the increased latency remains within 35\% of the baseline latency. Notably the latency of \sysname-EL3 even surpasses the baseline when the throughput increases.

\ignore{
The real-world applicaton overhead is much less than that of micro-benchmark, because the transitioning overhead only occupy a small proportion of total executing time. However, the worst overhead is 17$\%$ for the I/O intensive workloads like MongoDB. This is due to two main reasons. Firstly, because of the frequency MMIO and Interrupt emulation, the I/O intensive workload has much more \cVM exit and world switch than CPU intensive workload. In this case, the performance overhead introduced by micro operation which is shown in Table~\ref{fig:micro_benchmark} will be amplified for the I/O intensive workload. Secondly, because of the inaccessible  \virtio frontend shared memory, \sysname uses the \virtio SYNC module in \TMM to synchronize the \DMA buffer between the \cVM and host. However, the vanilla KVM could access the guest VM’s memory in default  \virtio mechanism directly. We expect the overheads for I/O intensive workloads could be significantly reduced by secure device passthrough in the future.
}

\section{Related Works}
\label{sec:related}


\para{TEE architectures}
All major processor manufactures announced the support of VM-based TEEs in their production roadmap, including AMD SEV~\cite{kaplan2016amd}, Intel TDX~\cite{tdx2020}, IBM PEF~\cite{hunt2021confidential} and Arm CCA~\cite{armcca}. Lee et al. propose Keystone~\cite{lee2020keystone}, an open and modular framework for building customizable TEEs. Similarly, CURE~\cite{bahmani2021cure} is a \TEE architecture with strongly-isolated and highly customizable enclaves, supporting a various types of enclave, ranging from enclaves in user space, over sub-space enclaves, to self-contained (multi-core) enclaves which include privileged software levels secure enclave-to-peripheral binding. 

\para{Hypervior-based TEEs}
SeKVM~\cite{li2021secure} and protected KVM (pKVM)~\cite{pkvm} offer a trusted core within the hypervisor to protect the confidentiality and integrity of \cVM data from untrusted host kernel. HyperEnclave~\cite{jia2022hyperenclave} employs a dedicated hypervisor that provides security isolation using commercial hardware virtualization technology, without relying on specific hardware features. In contrast, \sysname runs the \cvms in the TrustZone secure world, without trusting any components in the hypervisor.

\para{Re-purposing existing TEEs}
In addition to hardware-based \TEE proposals, efforts have been made to create extensible \TEE frameworks by leveraging existing \TEE solutions available on commodity CPUs. Examples of such frameworks include vTZ~\cite{hua2017vtz}, SecTEE~\cite{zhao2019sectee}, Sanctuary~\cite{brasser2019sanctuary}, vSGX~\cite{zhao2022vsgx}, and Twinvisor~\cite{li2021twinvisor}.
Among these works, Twinvisor is most closely related to our work. However, \sysname distinguishes itself from Twinvisor through several key design decisions. Notably, \sysname supports the CCA interface at the API level, allowing a CCA software stack to be built on it. Additionally, we have developed the entire software stack to support both \sysname and CCA. In contrast, Twinvisor only works with S-EL2 support, while \sysname offers compatibility with legacy hardware, regardless of S-EL2 support.
\section{Conclusion}
\label{sec:conclusion}

To enable the CCA model on legacy hardware, this paper presents \sysname, a virtualized CCA platform built upon existing TrustZone infrastructure. \sysname can be instantiated with the S-EL2 extension introduced in Arm V8.4, as well as on earlier platforms without S-EL2. Moreover, \sysname maintains strong compatibility with CCA at the API level, and we provide the necessary firmware and software stack to support \sysname.
We have implemented and evaluated \sysname on real Arm servers both with and without S-EL2 support. 



\balance
\bibliographystyle{plain}
\bibliography{ref}

\end{document}